%Paper: astro-ph/9506059
%From: burrows@cobalt.physics.arizona.edu
%Date: Wed, 7 Jun 1995 20:07:54 -0700

\magnification=\magstep1
\def\etal{{\it et.~al}}

\baselineskip=24truept

{\bf Probing Dark Matter}

{\bf by Adam Burrows and James Liebert}
\medskip

Recent novel observations
have probed the baryonic fraction of the galactic dark matter
that has eluded astronomers for decades.
Late in 1993, the MACHO$^1$ and EROS$^2$ collaborations announced in this
journal
the detection
of transient and achromatic brightenings of a handful of stars in the Large
Magellanic Cloud (LMC) that are best interpreted as gravitational
microlensing$^3$ by low-mass foreground objects (\underbar{MA}ssive
\underbar{C}ompact \underbar{H}alo \underbar{O}bjects, ``MACHOs'').
This tantalized astronomers, for it
implied that the population of cool, compact objects these lenses
represent could be the elusive dark matter of our galactic halo.
A year later in 1994, Sackett \etal\/$^4$ reported the discovery of a red
halo in the galaxy NGC 5907 that seems to follow the inferred radial
distribution
of its dark matter. This suggested that dwarf stars could constitute its
missing component.
Since NGC 5907 is similar to the Milky Way in type and radius, some surmised
that
the solution of the galactic dark matter problem was an abundance of ordinary
low-mass stars.

Now Bahcall \etal\/$^5$, using the Wide-Field Camera of the recently repaired
Hubble Space
Telescope, have dashed this hope. In a letter to the
Astrophysical Journal, they report the results of a deep pencil-beam search
in the V and I spectral bands for red dwarfs in our galaxy.
Surveying a high-latitude patch of the sky 4.4 square arcminutes
in area, Bahcall \etal\ find very few such stars and
conclude that red dwarfs above the stellar edge can
contribute no more than 6\% to the mass of our dark halo and no more than 15\%
to the mass of the
galactic disk.
One intriguing consequence of this observation is that if the microlenses are
not in the
LMC itself$\,^6$ and the halo is indeed made of
MACHOs, they are not stars above the hydrogen-burning limit, but
brown dwarfs below it.
However, if the MACHOs are not the dark matter, then the results of Bahcall
\etal\ imply that the missing galactic mass
has a particle-physics solution. Either way, the scientific community has
recently accelerated its search for the dominant
component of the galaxy.

What distinguishes the HST observations of Bahcall \etal\ is that they were
done from space with
unmatched angular resolution.  Resolutions of $\sim0.1$ arcseconds allow
astronomers to discriminate
between point dwarf stars and the extended galaxies that dominate a field
deeper than $\sim21$ magnitudes
in the visible.  Since competitive pencil-beam surveys are at least five
magnitudes deeper than this, it is generally
thought that one must be able to
separate stars from galaxies to obtain a credible red star census.
However, few extragalactic objects
intrude on the color range of the low mass Population~II stars (subdwarfs). In
studies of this population,
the star-galaxy separation problem is moot.  It is
appropriate, then, to ask how well the HST result agrees with
Pop~II studies made from the ground?

Dahn \etal\/$^7$ have recently estimated the luminosity function (LF)
of a kinematically-selected sample of Pop~II (visible
spheroid) stars in the solar neighborhood.
Most of the stars in their sample had trigonometric parallaxes (and, hence,
directly-measured distances),
a feature that deep
pencil-beam surveys lack.  The Dahn \etal\ LF peaks sharply near $M_V$ = 12
($M_I$ = 10) and turns downward towards an apparent terminus near $M_V$ =
14-14.5 ($M_I$ = 11).  They concluded that the subdwarfs from
the halo comprise only about 1/1000'th of the mass in stars in the solar
neighborhood -- approximately what Bahcall \etal\ derive from space.
If we extrapolate the Dahn \etal\ LF to the HST field and
assume that the Galactic density goes as $R^{-3.5}$ for
the visible spheroid, we predict what Bahcall \etal\ in fact saw: only a
handful of stars.
However, if this LF were applied to a baryonic ``dark halo''
with a local density of 0.009 solar masses per cubic parsec$^8$
and an $R^{-2}$ density
dependence, then upwards of 60 stars should have appeared in the
HST field (as Bahcall \etal\ point out).

Deep ground-based pencil-beam surveys
have pushed the CCD detector state-of-the-art to fainter
magnitudes, using telescopes larger in aperture than the HST
and covering larger areas of the sky.  Particularly important have
been the surveys of Tyson$^9$, Hu \etal\/$^{10}$, and Boeshaar, Tyson,
and Bernstein$^{11}$.
These workers probed larger volumes of space than Bahcall \etal\
and estimated
Pop~II low mass star
densities consistent with both the Bahcall \etal\ and Dahn \etal\ results.
The only LF inconsistent with these ground-based studies and the
HST study is that Richer
and Fahlman$^{12}$, whose LF is rising sharply down to the main
sequence limit.

The dearth of edge stars, either dwarfs or low-metallicity subdwarfs,
allows us to conclude with some certainty that neither red dwarfs nor
subdwarfs can be a major mass fraction of any component of the galaxy.
We are left with a classic mystery: we think that there are compact
microlenses between us and the LMC, but we can not see them directly
with our best cameras.  Furthermore, if they are old brown dwarfs,
we can not explain why they were formed as a distinct population that
is not a simple extrapolation of the stars that we do see.

These novel surveys demonstrate just how great has been the recent
improvement in search technology. Deep pencil-beam surveys have the
potential to provide new and important data on the nature of the
galactic halo (and what it can not be) that will complement those now
being obtained by the microlensing searches sensitive only to
gravitational mass.  All too often, discussions of the halo dark
matter have resembled medieval discourses on the Aristotelean
quintessence or the angelic population of the empyrean. Astronomers
seemed to be involved in bootless shadow boxing
with a Nature jealous of its secrets. With the recent deep
photometric and microlensing surveys, we may finally be learning
something of substance concerning the dominant constituents of our
galaxy and, perhaps, the universe.  \bigskip

{\it Adam Burrows is in the Departments of Physics and Astronomy and is
chairman of the Theoretical Astrophysics Program of the University of
Arizona, Tucson, Arizona 85721 USA.  James Liebert is affiliated with the
Department of Astronomy and Steward Observatory at the same
institution.}

\bigskip

\centerline{\bf References}
\bigskip
\item{1.} Alcock, C.~\etal\ (the MACHO collaboration) {\it Nature} {\bf 365},
621--623 (1993).
\item{2.} Aubourg, E.~\etal\ (the EROS collaboration) {\it Nature} {\bf 365},
623--625 (1993).
\item{3.} Paczynski, B.\ {\it Astrophys.\ J.} {\bf 304}, 1--5 (1986).
\item{4.} Sackett, P. ~\etal\ {\it Nature} {\bf 370}, 441 (1994).
\item{5.} Bahcall, J. N. ~\etal\ \ {\it Astrophys.\ J.} {\bf 435}, L51--L54
(1994).
\item{6.} Sahu, K. C. \ {\it Nature} {\bf 370}, 275 (1994).
\item{7.} Dahn, C., Liebert, J., Harris, H., \& Guetter, H. C. \
to appear in {\bf An ESO Workshop on: The Bottom of the Main Sequence
and Beyond}, ed. C.G. Tinney, Berlin: Springer-Verlag, in press (1994).
\item{8.} Bahcall, J.N., Schmidt, M., \& Soneira, R.M. {\it Astrophys.\ J.}{\bf
265}, 730 (1983).
\item{9.} Tyson, J. A.\ {\it Astron.\ J.} {\bf 96}, 1--23 (1988).
\item{10.} Hu, E.~\etal\ {\it Nature} {\bf 371}, 493 (1994).
\item{11.} Boeshaar, P., Tyson, J. A., \& Bernstein, G. M. to appear in {\bf
Dark Matter}, the
5'th Maryland Astrophysics Conference, Oct. 1994.
\item{12.} Richer, H. B. \& Fahlman, G. G. {\it Nature} {\bf 358}, 353 (1992).
\bye